# General Criterion for Controllable Conformational Transitions of Single and Double Stranded DNA

Haim Diamant[*] & David Andelman[†]

[*]*James Franck Institute, University of Chicago, Chicago, Illinois 60637, USA*

[†]*School of Physics and Astronomy, Raymond and Beverly Sackler Faculty of Exact Sciences, Tel Aviv University, 69978 Tel Aviv, Israel*

**Chain-like macromolecules in solution, whether biological or synthetic, transform from a spatially extended conformation to a compact one upon change of temperature or solvent qualities. This sharp transition plays a key role in various phenomena, including DNA condensation[1], protein folding[2,3], and the behaviour of polymer solutions[4]. In biological processes such as DNA condensation the collapse is sensitively induced by a small amount of added molecules. Here we derive a general criterion for the effect of such agents on conformational transitions. We find two different scenarios depending on chain stiffness. If the persistence length —the characteristic distance along which the chain retains its direction— is smaller than the range of attractive correlations induced by the agent (typically up to several nanometres), the chain contracts gradually. Stiffer chains undergo sharp collapse. We thereby suggest that the enhanced rigidity of double-stranded DNA as compared to the single strand is a prerequisite for sharp, controllable conformational transitions.**

When material scientists wish to change the conformation of dissolved polymers, thereby controlling the phase and flow behaviour of the solution[4], they tune overall parameters such as temperature, pH or salinity. In biological systems a similar goal is achieved by introducing a small quantity of condensing agents, such as short polyamines (spermine and spermidine) in the case of DNA condensation[1,5,6]. Studies have shown that DNA collapse can be induced by other, non-specific agents, *e.g.* inorganic multivalent ions[7] and ionic surfactants[8,9]. Apart from the basic interest in DNA condensation, such mixtures are important for potential gene delivery applications, where the DNA is 'shielded' by oppositely charged molecules to help it penetrate the cell[1,10].

Theoretical studies have been focused on the complex electrostatics among chain groups and the surrounding ions[10], in particular the perplexing emergence of attraction between like-charge DNA segments. While electrostatics evidently plays a central role (all known condensing agents are charged) there are other factors to consider. For example, monovalent ionic surfactants can condense DNA[8,9], whereas simpler ions must be multivalent[10]. Beyond mere charge, the key feature of a successful agent seems to be its ability to co-operatively associate with the chain, thereby inducing strong attractive correlations. Indeed, all of the aforementioned agents can be viewed as molecular 'clips' —having associated with a monomer, they attract other monomers and agent molecules to the same region. Approaching the problem from a new direction, we assume that the added agent induces attractive correlations in the chain and ask what is required of the biopolymer to ensure sharp collapse.

We consider a DNA molecule, either single-stranded (ssDNA) or double-stranded (dsDNA), as an isolated long chain in aqueous solution. The chain stiffness is characterised by the *persistence length*, $l_p$. dsDNA is very stiff, having $l_p$ in the range of 50–100 nm[11,12], whereas ssDNA is much more flexible, with $l_p$ estimated between a fraction of nanometre and 2–3 nm[13–15]. Chain sections shorter than $l_p$ have a one-dimensional character, like stiff rods, while at scales larger than $l_p$ the chain is easily deformed and assumes a swollen-coil conformation.

Four assumptions underlie our analysis. (i) The chain is taken to be very long compared to its $l_p$. (ii) As is valid for long chains, we assume that, in the absence of condensing agent, there is a sharp collapse transition at a certain value of control parameter (*e.g.* temperature or pH). This transition is represented schematically by the point $T^*$ on the $T$ axis in Fig. 1. We consider the effect of added agents as modifying this *pre-existing* transition. (iii) The collapse is assumed to be initiated isotropically, similar to that of synthetic polymers. Most biopolymers actually fold into ordered, anisotropic structures. Yet it is known, at least for proteins[2,3] and RNA[16], that the chains initially collapse into an intermediate isotropic state. (iv) The range $\xi$ of the agent-induced attraction is assumed to be larger than the monomer size and bare monomer–monomer interactions. The last requirement is only marginally fulfilled in practice. Yet this should not affect the main finding —that the qualitative collapse behaviour is determined by an interplay between $\xi$ and $l_p$.

Consider now the addition of a small amount $\Phi$ of condensing agent. Once these molecules associate with the chain, two scenarios are possible. First, the transition point $T^*$ may turn into a sharp line, $T^* = T^*(\Phi)$, as schematically shown by the solid line in Fig. 1. The other possibility is that the point broadens into a finite *region* in the $(T,\Phi)$ plane, as is schematically represented by the grey area between dashed lines; this means that the collapse has turned into gradual contraction over a finite concentration range. Our aim is to find the criterion that ensures the former scenario, *i.e.* sharp collapse as a function of $\Phi$. This scenario involves a considerable loss of entropy; if possible, the chain will avoid it by a gradual response. Let us assume, therefore, that the gradual-collapse scenario holds over a range of $\Phi$, and check when this assumption can be consistently satisfied.

The effect of the condensing agent may be represented by attraction induced between any two monomers. This interaction takes the general form $u(r) = -Af(r/\xi)$, where $r$ is the distance between two monomers, $A$ a coupling parameter (assumed weak compared to the thermal energy $k_B T$), $\xi$ the correlation length, and $f(r/\xi)$ a function that decays fast to zero for $r>\xi$. The interaction parameters $A$ and $\xi$ are tuned by changing Φ. The overall effect can be described by a reduction in the excluded-volume parameter of the chain, $v(T,\Phi) = v_0(T) - \delta v(T,\Phi) < v_0$, where $v_0$ is the excluded-volume parameter of the bare chain, and

$$\delta v(T,\Phi) \cong -\int d\vec{r} u(r)/k_B T \sim A\xi^3 . \quad (1)$$

As long as $v>0$, the overall monomer–monomer interaction is repulsive and the chain remains swollen[4].

Gradual contraction implies that the collapse does not involve all the monomers. The chain can be envisioned as divided into subunits, as drawn in Fig. 2, such that the different interactions dominate on different length scales —a section of monomers contained in a subunit has the properties of the bare chain, whereas over larger distances the induced attraction dominates. (Similar ideas of chain 'rescaling' are often used in polymer physics[4].) As the control parameters are changed within the gradual-collapse region, the sub-division is adapted so as to keep the competing interactions just balanced, *i.e.* the *rescaled* chain of subunits remains just at a collapse point[17–20].

Imagine, therefore, that the chain is divided into subunits of size $b$, each containing $g$ monomers (Fig. 2). To obtain the behaviour of the rescaled chain we need to calculate the effective interaction $U(r)$ acting between two subunits. This consists of strong repulsion for distances smaller than $b$, $U(r<b) \to \infty$, and attraction for distances larger than $b$ coming from the integrated interaction of $g^2$ pairs of monomers in the two subunits, $U(r>b) = g^2 u(r)$. Balance between the competing interactions requires that the rescaled excluded-volume parameter vanish, $V = \int d\vec{r}\left[1 - e^{-U(r)/k_B T}\right] = 0$. Substituting the expressions for $U(r)$ and $u(r)$ we obtain an implicit relation for the chain sub-division parameters, $b$ and $g$,

$$\int_b^\infty r^2 dr\left[1 - e^{g^2 A f(r/\xi)/k_B T}\right] = const. \qquad (2)$$

The ability of the chain to adjust its sub-division requires that Eq. (2) be satisfied for *any* value of Φ within the cross-over region. This can be achieved only if both $[g(\Phi)]^2 A(\Phi)$ and $b(\Phi)/\xi(\Phi)$ remain constant. Thus, the subunit size $b$ changes proportionally with $\xi$. In addition, $g$ is generally related to $b$ by a scaling relation, $g \sim b^D \sim \xi^D$, where $D$ is the subunit (fractal) dimensionality. Then, substituting Eq. (1), we obtain

$$g(\Phi) \sim [\delta v(\Phi)]^{-D/(2D-3)}, \qquad b(\Phi) \sim [\delta v(\Phi)]^{-1/(2D-3)}. \qquad (3)$$

For the gradual scenario to be self-consistent, the subunits should shrink upon agent addition. For example, when Φ is very low the entire chain is a single unperturbed unit, whereas for high enough Φ the chain is completely collapsed and each subunit contains a single monomer. Since the overall effect of the agent, $\delta v$, must increase with Φ, the self-consistency criterion is, according to Eq. (3), $D>3/2$. The subunit dimensionality is not a continuous parameter; for $b \sim \xi < l_p$ the subunit is rod-like, *i.e.* $D=1$, while for $b > l_p$ it has the dimensionality of self-avoiding random walks, $D \approx 1.7$ [4]. Hence, in practical terms, the criterion just found is

$$l_p < \xi \quad \text{gradual cross-over,}$$
$$l_p > \xi \quad \text{sharp transition.} \qquad (4)$$

The condition $D>3/2$ is a manifestation of a more general principle from polymer physics known as the *des Cloizeaux criterion*[21], which suggests a geometric interpretation of our result. In order for the interactions to be relevant and broaden the pre-existing transition, a pair of subunits must not be 'mutually transparent', *i.e.* the probability of their coming into close proximity must be non-vanishing[22]. This requires that the sum of fractal dimensions of the two interacting objects exceed 3, the dimension of the embedding space, hence $D>3/2$. We now demonstrate the application of criterion (4) in two types of systems corresponding to the two different scenarios.

An example for the sharp case is provided by solutions of DNA and surfactants investigated in the past years[8,9]. As discussed above, dsDNA is stiff on length scales up to 50 nm, thus $l_p>\xi$. Indeed, sharp collapse was convincingly demonstrated by fluorescence microscopy[8,9], as reproduced in Fig. 3. Similar transitions were found upon addition of short polyamines[5,6] and multivalent salts[7].

By contrast, ssDNA is a flexible chain characterised by rather short persistence lengths, $l_p<\xi$. Hence, upon addition of surfactant or polyamine, its folding should be gradual. This specific prediction for ssDNA still awaits experimental confirmation. (Similar conclusions should hold for double-stranded *vs.* single-stranded RNA[23].) Our general prediction for flexible chains, nonetheless, agrees with numerous experiments on solutions of synthetic polymers and surfactants widely used in industry[24]. Polymer–surfactant interaction is typically studied upon increasing surfactant concentration at constant temperature. Such a procedure is represented by the vertical dotted arrow in Fig. 1. A flexible polymer such as polyethylene glycol has $l_p$ of less than a nanometre and should undergo gradual contraction. Experiments find, indeed, a concentration *range* within which the association takes place. At a certain concentration $\Phi_1$ (so-called *critical aggregation concentration*) the polymer and surfactant start associating, and only at a higher concentration, $\Phi_2$, do the individual polymers become compact and usually precipitate[24]. There is substantial evidence from neutron scattering[25], light scattering[26], and rheology[27] that the polymer gradually contracts above $\Phi_1$, but does not sharply collapse.

If dsDNA had a short persistence length of less than 10 nm, the sharp condensation would be replaced by a less controllable, smooth change. Hence, the enhanced stiffness of dsDNA as a result of the double-helical structure (and possibly of other long biopolymers) appears to be a necessary condition for sharp, tuneable conformational transitions.

We are grateful to T. Witten for many helpful suggestions. We benefited from discussions and correspondence with R. Bar-Ziv, T. Garel, B. Harris, M. Kozlov, S. Leikin, S. Marčelja, A. Meller, H. Orland, M. Schwartz, V. Sergeyev and K. Yoshikawa. This work was supported by the Israel Science Foundation (Centers of Excellence Program), the Israel–US Binational Science Foundation, the National Science Foundation, the NSF MRSEC program, and the American Lung Association.

**Correspondence and requests for materials should be addressed to D.A. (e-mail: andelman@post.tau.ac.il).**

Fig. 1: Schematic diagram showing the dependence of the collapse transition on a control parameter *T* and added agent concentration $\Phi$. In one scenario the transition remains sharp upon agent addition (solid line), whereas in the other the transition is broadened into a gradual cross-over (grey area between dashed lines). The dotted arrow represents a common procedure of adding agent at constant *T*.

Fig. 2: Schematic view of a part of chain undergoing gradual contraction due to added agent (represented by solid spheres). The chain can be divided into sub-units (shown enclosed by fictitious shells) whose size is self-adjusted with agent concentration, such that the rescaled chain is kept collapsed. This scenario is relevant to flexible chains such as ssDNA.

Fig. 3: Sharp collapse of dsDNA upon surfactant addition, as visualised by fluorescence microscopy. a) Swollen state in presence of surfactant $D_{18}DAB$ at concentration $\Phi$=1.6x10$^{-9}$ M. The bar represents 10 microns. b) Sharp appearance of collapsed chains at $\Phi$=1.0x10$^{-8}$ M in coexistence with swollen ones. c) Above $\Phi$=4.0x10$^{-6}$ M all chains are collapsed. Note the extremely low concentration required to induce collapse; it is over 100 times lower than the onset of pure surfactant self-association. (Adapted from Ref. 9 —reproduced by permission of the Royal Society of Chemistry.)

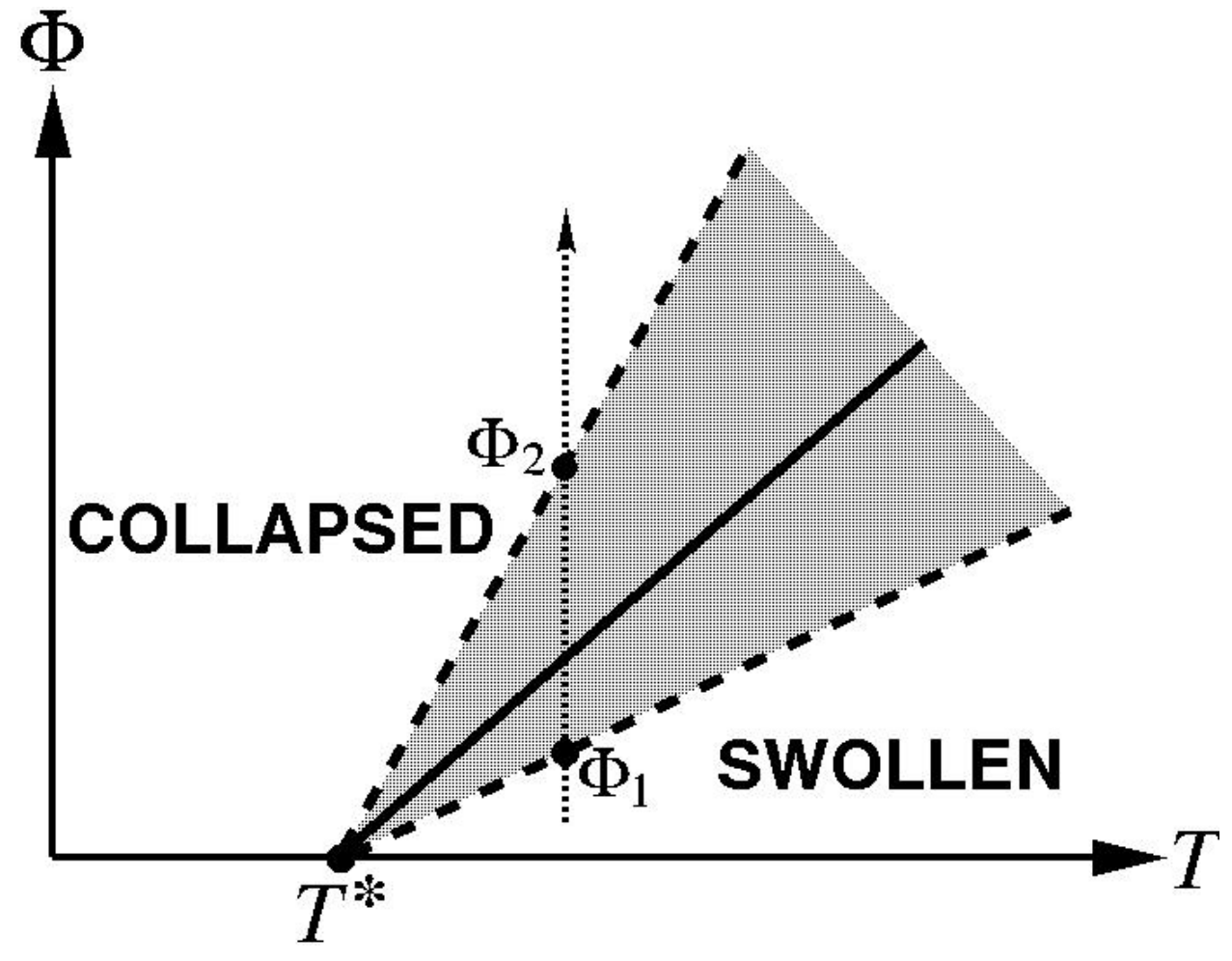

Fig. 1

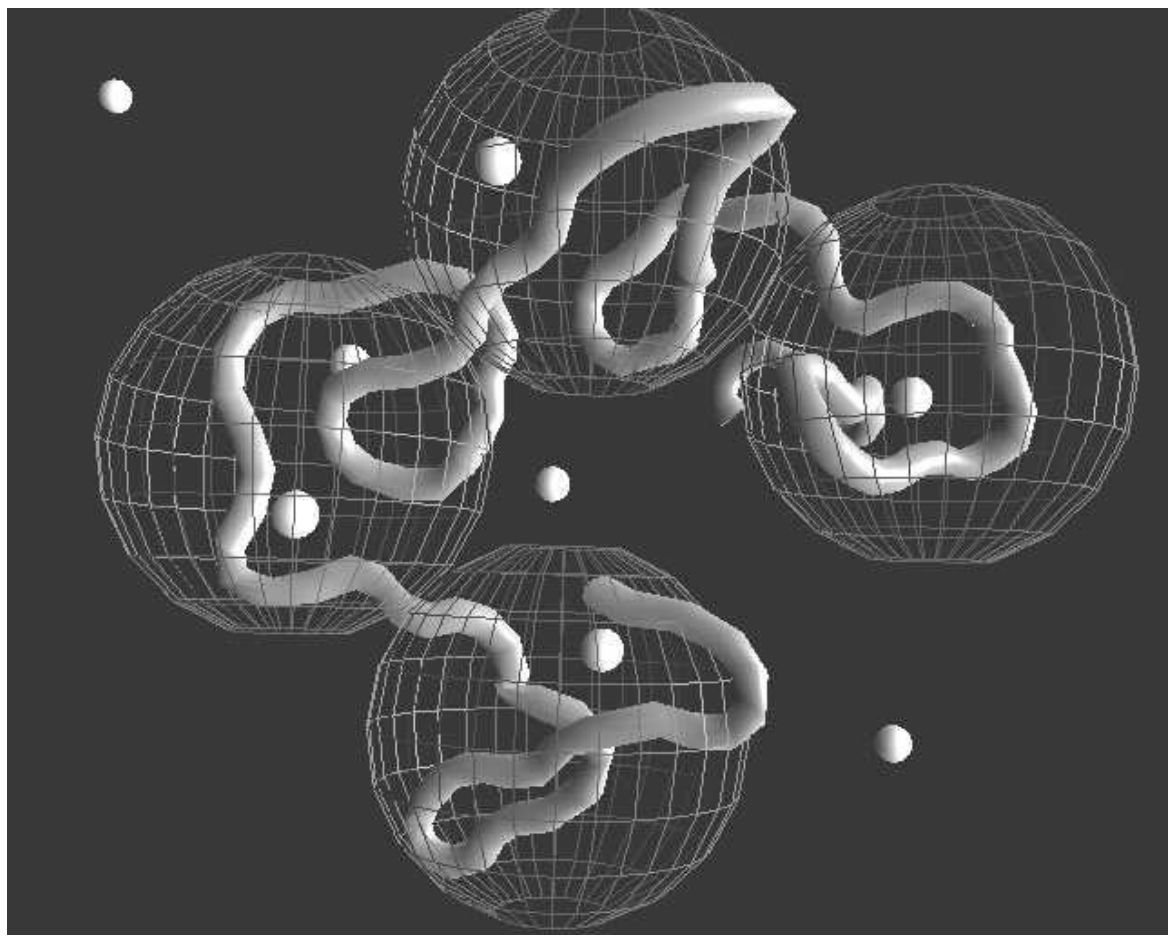

Fig. 2

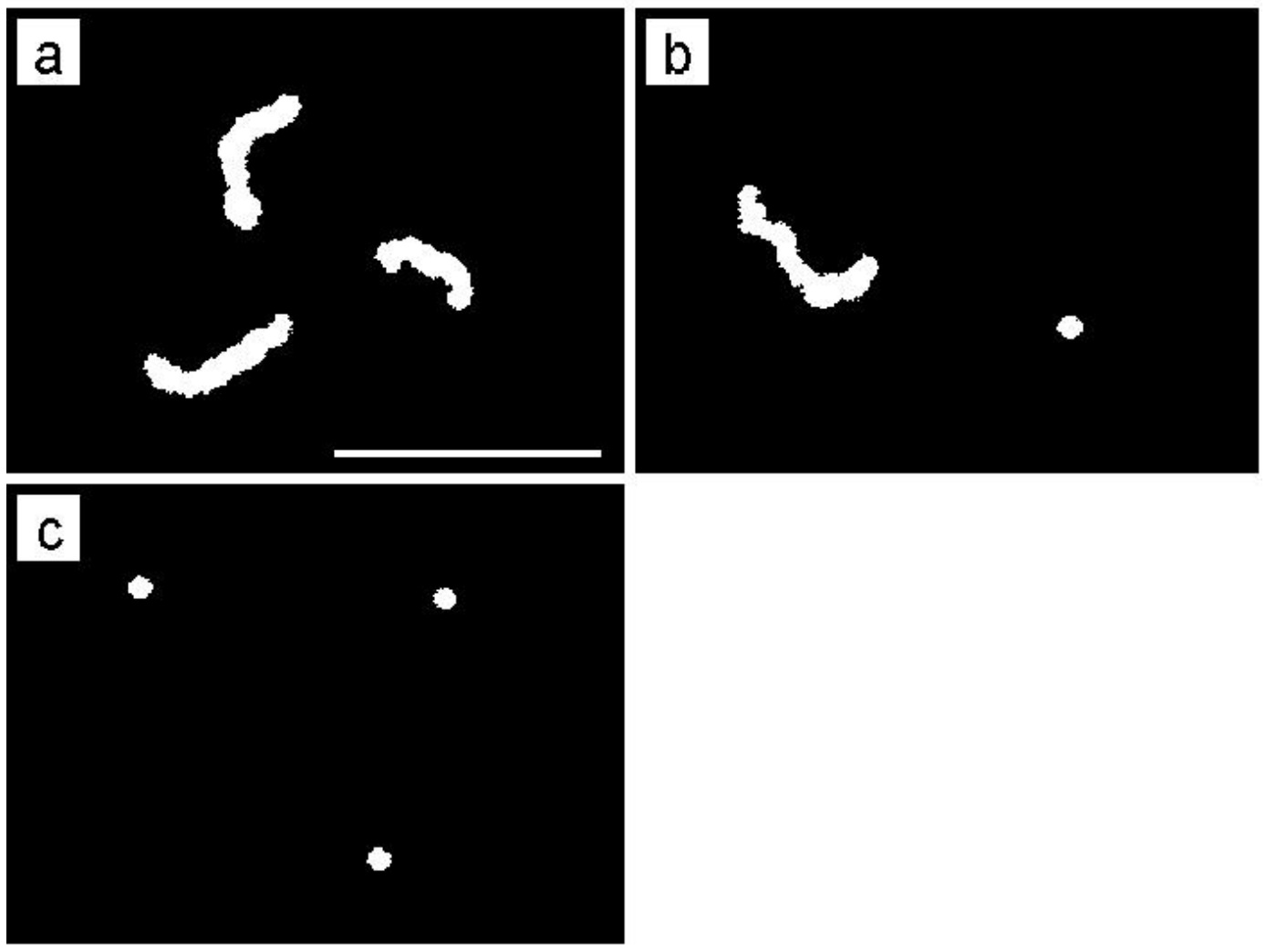

Fig. 3